%Paper: hep-th/9207013
%From: "Jose M. Figueroa-O'Farrill" <figueroa@pib1.physik.uni-bonn.de>
%Date: Fri, 03 Jul 1992 23:07:20 MEZ

%%%%%%%%%%%%%%%%%%%%%%%%%%%%%%%%%%%%%%%%%%%%%%%%%%%%%%%%%%%
%                                                         %
%  On the two-boson picture of the KP hierarchy           %
%                                                         %
%              (8 pages, Plain TeX)                       %
%                                                         %
%  Jose M. Figueroa-O'Farrill, Javier Mas, Eduardo Ramos  %
%      BONN-HE-92-17, KUL-TF-92/26, US-FT-4/92            %
%                                                         %
%%%%%%%%%%%%%%%%%%%%%%%%%%%%%%%%%%%%%%%%%%%%%%%%%%%%%%%%%%%
%
%  This is a stripped down version of the macros for submission
%  to the hep-th bulletin board. Last revised: June 10th, 1992.
%
%
\nonstopmode
\catcode`\@=11 % this allows for tricky names
%
%%%%%%%%%%%%%%%%%%%%%%%%%%%%%%%%%%%%%%%%%%%%%%%%%%%%%%%%%%%%%%%%%
%
%  First some font definitions
%
\font\seventeenrm=cmr17

\font\twelverm=cmr12
\font\ninerm=cmr9
\font\sixrm=cmr6

\font\seventeenbf=cmbx12 at 17pt
\font\fourteenbf=cmbx12 at 14pt
\font\twelvebf=cmbx12
\font\ninebf=cmbx9
\font\sixbf=cmbx6

\font\seventeeni=cmmi12 at 17pt             \skewchar\seventeeni='177
\font\fourteeni=cmmi12 at 14pt              \skewchar\fourteeni='177
\font\twelvei=cmmi12                        \skewchar\twelvei='177
\font\ninei=cmmi9                           \skewchar\ninei='177
\font\sixi=cmmi6                            \skewchar\sixi='177

\font\seventeensy=cmsy10 scaled\magstep3    \skewchar\seventeensy='60
\font\fourteensy=cmsy10 scaled\magstep2     \skewchar\fourteensy='60
\font\twelvesy=cmsy10 at 12pt               \skewchar\twelvesy='60
\font\ninesy=cmsy9                          \skewchar\ninesy='60
\font\sixsy=cmsy6                           \skewchar\sixsy='60

\font\seventeenex=cmex10 scaled\magstep3
\font\fourteenex=cmex10 scaled\magstep2
\font\twelveex=cmex10 at 12pt

%\font\ninex=cmex9
\font\ninex=cmex10 at 9pt
%\font\sevenex=cmex7
\font\sevenex=cmex10 at 9pt
%\font\sixex=cmex7 at 6pt
\font\sixex=cmex10 at 6pt
%\font\fivex=cmex7 at 5pt
\font\fivex=cmex10 at 5pt

\font\seventeensl=cmsl10 scaled\magstep3
\font\fourteensl=cmsl10 scaled\magstep2
\font\twelvesl=cmsl10 scaled\magstep1
\font\ninesl=cmsl10 at 9pt
\font\sevensl=cmsl10 at 7pt
\font\sixsl=cmsl10 at 6pt
\font\fivesl=cmsl10 at 5pt

\font\seventeenit=cmti12 scaled\magstep2
\font\fourteenit=cmti12 scaled\magstep1
\font\twelveit=cmti12

\font\seventeentt=cmtt12 scaled\magstep2
\font\fourteentt=cmtt12 scaled\magstep1
\font\twelvett=cmtt12

\font\seventeencp=cmcsc10 scaled\magstep3
\font\fourteencp=cmcsc10 scaled\magstep2
\font\twelvecp=cmcsc10 scaled\magstep1
\font\tencp=cmcsc10
%\font\eightcp=cmcsc8

\newfam\cpfam

\font\seventeenss=cmss17
\font\fourteenss=cmss12 at 14pt
\font\twelvess=cmss12
\font\tenss=cmss10
\font\niness=cmss9

\font\sevenss=cmss8 at 7pt
\font\sixss=cmss8 at 6pt
\font\fivess=cmss8 at 5pt
\newfam\ssfam
\newdimen\b@gheight             \b@gheight=12pt
\newcount\f@ntkey               \f@ntkey=0
\def\f@m{\afterassignment\samef@nt\f@ntkey=}
\def\samef@nt{\fam=\f@ntkey \the\textfont\f@ntkey\relax}
\def\rm{\f@m0 }
\def\mit{\f@m1 }         
\def\cal{\f@m2 }
\def\it{\f@m\itfam}
\def\sl{\f@m\slfam}
\def\bf{\f@m\bffam}
\def\tt{\f@m\ttfam}
\def\ssf{\f@m\ssfam}
\def\caps{\f@m\cpfam}
\def\seventeenpoint{\relax
    \textfont0=\seventeenrm          \scriptfont0=\twelverm
      \scriptscriptfont0=\ninerm
    \textfont1=\seventeeni           \scriptfont1=\twelvei
      \scriptscriptfont1=\ninei
    \textfont2=\seventeensy          \scriptfont2=\twelvesy
      \scriptscriptfont2=\ninesy
    \textfont3=\seventeenex          \scriptfont3=\twelveex
      \scriptscriptfont3=\ninex
    \textfont\itfam=\seventeenit    %\scriptfont\itfam=\twelveit
    \textfont\slfam=\seventeensl    %\scriptfont\slfam=\twelvesl
      \scriptscriptfont\slfam=\ninesl
    \textfont\bffam=\seventeenbf     \scriptfont\bffam=\twelvebf
      \scriptscriptfont\bffam=\ninebf
    \textfont\ttfam=\seventeentt
    \textfont\cpfam=\seventeencp
    \textfont\ssfam=\seventeenss     \scriptfont\ssfam=\twelvess
      \scriptscriptfont\ssfam=\niness
    \samef@nt
    \b@gheight=17pt
    \setbox\strutbox=\hbox{\vrule height 0.85\b@gheight
                                depth 0.35\b@gheight width\z@ }}
\def\fourteenpoint{\relax
    \textfont0=\fourteencp          \scriptfont0=\tenrm
      \scriptscriptfont0=\sevenrm
    \textfont1=\fourteeni           \scriptfont1=\teni
      \scriptscriptfont1=\seveni
    \textfont2=\fourteensy          \scriptfont2=\tensy
      \scriptscriptfont2=\sevensy
    \textfont3=\fourteenex          \scriptfont3=\twelveex
      \scriptscriptfont3=\tenex
    \textfont\itfam=\fourteenit     \scriptfont\itfam=\tenit
    \textfont\slfam=\fourteensl     \scriptfont\slfam=\tensl
      \scriptscriptfont\slfam=\sevensl
    \textfont\bffam=\fourteenbf     \scriptfont\bffam=\tenbf
      \scriptscriptfont\bffam=\sevenbf
    \textfont\ttfam=\fourteentt
    \textfont\cpfam=\fourteencp
    \textfont\ssfam=\fourteenss     \scriptfont\ssfam=\tenss
      \scriptscriptfont\ssfam=\sevenss
    \samef@nt
    \b@gheight=14pt
    \setbox\strutbox=\hbox{\vrule height 0.85\b@gheight
                                depth 0.35\b@gheight width\z@ }}
\def\twelvepoint{\relax
    \textfont0=\twelverm          \scriptfont0=\ninerm
      \scriptscriptfont0=\sixrm
    \textfont1=\twelvei           \scriptfont1=\ninei
      \scriptscriptfont1=\sixi
    \textfont2=\twelvesy           \scriptfont2=\ninesy
      \scriptscriptfont2=\sixsy
    \textfont3=\twelveex          \scriptfont3=\ninex
      \scriptscriptfont3=\sixex
    \textfont\itfam=\twelveit    %\scriptfont\itfam=\nineit
    \textfont\slfam=\twelvesl    %\scriptfont\slfam=\ninesl
      \scriptscriptfont\slfam=\sixsl
    \textfont\bffam=\twelvebf     \scriptfont\bffam=\ninebf
      \scriptscriptfont\bffam=\sixbf
    \textfont\ttfam=\twelvett
    \textfont\cpfam=\twelvecp
    \textfont\ssfam=\twelvess     \scriptfont\ssfam=\niness
      \scriptscriptfont\ssfam=\sixss
    \samef@nt
    \b@gheight=12pt
    \setbox\strutbox=\hbox{\vrule height 0.85\b@gheight
                                depth 0.35\b@gheight width\z@ }}
\def\tenpoint{\relax
    \textfont0=\tenrm          \scriptfont0=\sevenrm
      \scriptscriptfont0=\fiverm
    \textfont1=\teni           \scriptfont1=\seveni
      \scriptscriptfont1=\fivei
    \textfont2=\tensy          \scriptfont2=\sevensy
      \scriptscriptfont2=\fivesy
    \textfont3=\tenex          \scriptfont3=\sevenex
      \scriptscriptfont3=\fivex
    \textfont\itfam=\tenit     \scriptfont\itfam=\seveni
    \textfont\slfam=\tensl     \scriptfont\slfam=\sevensl
      \scriptscriptfont\slfam=\fivesl
    \textfont\bffam=\tenbf     \scriptfont\bffam=\sevenbf
      \scriptscriptfont\bffam=\fivebf
    \textfont\ttfam=\tentt
    \textfont\cpfam=\tencp
    \textfont\ssfam=\tenss     \scriptfont\ssfam=\sevenss
      \scriptscriptfont\ssfam=\fivess
    \samef@nt
    \b@gheight=10pt
    \setbox\strutbox=\hbox{\vrule height 0.85\b@gheight
                                depth 0.35\b@gheight width\z@ }}
%
%%%%%%%%%%%%%%%%%%%%%%%%%%%%%%%%%%%%%%%%%%%%%%%%%%%%%%%%%%%%%%%%%%%%%%%%
%
%   Next, I define basic spacing parameters.
%
\normalbaselineskip = 15pt plus 0.2pt minus 0.1pt %was 20pt ...
\normallineskip = 1.5pt plus 0.1pt minus 0.1pt
\normallineskiplimit = 1.5pt
\newskip\normaldisplayskip
\normaldisplayskip = 15pt plus 5pt minus 10pt %was 20pt ...
\newskip\normaldispshortskip
\normaldispshortskip = 6pt plus 5pt
\newskip\normalparskip
\normalparskip = 6pt plus 2pt minus 1pt
\newskip\skipregister
\skipregister = 5pt plus 2pt minus 1.5pt
\newif\ifsingl@    \newif\ifdoubl@
\newif\iftwelv@    \twelv@true
\def\singlespace{\singl@true\doubl@false\spaces@t}
\def\doublespace{\singl@false\doubl@true\spaces@t}
\def\normalspace{\singl@false\doubl@false\spaces@t}
\def\Tenpoint{\tenpoint\twelv@false\spaces@t}
\def\Twelvepoint{\twelvepoint\twelv@true\spaces@t}
\def\spaces@t{\relax
      \iftwelv@ \ifsingl@\subspaces@t3:4;\else\subspaces@t1:1;\fi
       \else \ifsingl@\subspaces@t3:5;\else\subspaces@t4:5;\fi \fi
      \ifdoubl@ \multiply\baselineskip by 5
         \divide\baselineskip by 4 \fi }
\def\subspaces@t#1:#2;{
      \baselineskip = \normalbaselineskip
      \multiply\baselineskip by #1 \divide\baselineskip by #2
      \lineskip = \normallineskip
      \multiply\lineskip by #1 \divide\lineskip by #2
      \lineskiplimit = \normallineskiplimit
      \multiply\lineskiplimit by #1 \divide\lineskiplimit by #2
      \parskip = \normalparskip
      \multiply\parskip by #1 \divide\parskip by #2
      \abovedisplayskip = \normaldisplayskip
      \multiply\abovedisplayskip by #1 \divide\abovedisplayskip by #2
      \belowdisplayskip = \abovedisplayskip
      \abovedisplayshortskip = \normaldispshortskip
      \multiply\abovedisplayshortskip by #1
        \divide\abovedisplayshortskip by #2
      \belowdisplayshortskip = \abovedisplayshortskip
      \advance\belowdisplayshortskip by \belowdisplayskip
      \divide\belowdisplayshortskip by 2
      \smallskipamount = \skipregister
      \multiply\smallskipamount by #1 \divide\smallskipamount by #2
      \medskipamount = \smallskipamount \multiply\medskipamount by 2
      \bigskipamount = \smallskipamount \multiply\bigskipamount by 4 }
\def\normalbaselines{ \baselineskip=\normalbaselineskip
   \lineskip=\normallineskip \lineskiplimit=\normallineskip
   \iftwelv@\else \multiply\baselineskip by 4 \divide\baselineskip by 5
     \multiply\lineskiplimit by 4 \divide\lineskiplimit by 5
     \multiply\lineskip by 4 \divide\lineskip by 5 \fi }
\Twelvepoint  % That's the default
\interlinepenalty=50
\interfootnotelinepenalty=5000
\predisplaypenalty=9000
\postdisplaypenalty=500
\hfuzz=1pt
\vfuzz=0.2pt
\dimen\footins=24 truecm % 8 truein in SB
\hoffset=10.5truemm % 0 in SB, 6.5mm in Leuven
\voffset=-8.5 truemm % 0in in SB, 5 truemm in Leuven
%
%%%%%%%%%%%%%%%%%%%%%%%%%%%%%%%%%%%%%%%%%%%%%%%%%%%%%%%%%%%%%%%%%%
%
% Now some output macros
%
%%%%%%%%%%%%%%%%%%%%%%%%%%%%%%%%%%%%%%%%%%%
%
% footnote numbering macros
%
%
\def\footnote#1{\edef\@sf{\spacefactor\the\spacefactor}#1\@sf
      \insert\footins\bgroup\singl@true\doubl@false\Tenpoint
      \interlinepenalty=\interfootnotelinepenalty \let\par=\endgraf
        \leftskip=\z@skip \rightskip=\z@skip
        \splittopskip=10pt plus 1pt minus 1pt \floatingpenalty=20000
        \smallskip\item{#1}\bgroup\strut\aftergroup\@foot\let\next}
\skip\footins=\bigskipamount % space added when footnote is present
\dimen\footins=24truecm % maximum footnotes per page (8 truein in USA)
\newcount\fnotenumber
\def\clearfnotenumber{\fnotenumber=0}
\def\fnote{\advance\fnotenumber by1 \footnote{$^{\the\fnotenumber}$}}
\clearfnotenumber
%
% section and appendix macros
%
\newcount\secnumber
\newcount\appnumber
\newif\ifs@c % this is true if within a section as opposed to an appendix
\newif\ifs@cd % this is true if the article is being section'd
\s@cdtrue % this is the default
\def\unsectioned{\s@cdfalse\let\section=\subsection}
\def\clearappnumber{\appnumber=64}
\def\clearsecnumber{\secnumber=0}
\newskip\sectionskip         \sectionskip=\medskipamount
\newskip\headskip            \headskip=8pt plus 3pt minus 3pt
\newdimen\sectionminspace    \sectionminspace=10pc
\newdimen\referenceminspace  \referenceminspace=25pc
\def\Titlestyle#1{\par\begingroup \interlinepenalty=9999
     \leftskip=0.02\hsize plus 0.23\hsize minus 0.02\hsize
     \rightskip=\leftskip \parfillskip=0pt
     \advance\baselineskip by 0.5\baselineskip%this is a test...
     \hyphenpenalty=9000 \exhyphenpenalty=9000
     \tolerance=9999 \pretolerance=9000
     \spaceskip=0.333em \xspaceskip=0.5em
     \seventeenpoint
  \noindent #1\par\endgroup }
\def\titlestyle#1{\par\begingroup \interlinepenalty=9999
     \leftskip=0.02\hsize plus 0.23\hsize minus 0.02\hsize
     \rightskip=\leftskip \parfillskip=0pt
     \hyphenpenalty=9000 \exhyphenpenalty=9000
     \tolerance=9999 \pretolerance=9000
     \spaceskip=0.333em \xspaceskip=0.5em
     \fourteenpoint
   \noindent #1\par\endgroup }%  the \Npoint only takes care of spacing.
%                                a font is always specified when calling this
%                                macro.  In computers with little room for
%                                character-size data it is convenient to % out
%                                all the font definitions from the \Npoint
%                                macros.
%
\def\spacecheck#1{\dimen@=\pagegoal\advance\dimen@ by -\pagetotal
   \ifdim\dimen@<#1 \ifdim\dimen@>0pt \vfil\break \fi\fi}
\def\section#1{\cleareqnumber \s@ctrue \global\advance\secnumber by1
   \message{Section \the\secnumber: #1}
   \par \ifnum\the\lastpenalty=30000\else
   \penalty-200\vskip\sectionskip \spacecheck\sectionminspace\fi
   \noindent {\caps\enspace\S\the\secnumber\quad #1}\par
   \nobreak\vskip\headskip \penalty 30000 }
\def\subsection#1{\par
   \ifnum\the\lastpenalty=30000\else \penalty-100\smallskip
   \spacecheck\sectionminspace\fi
   \noindent\undertext{#1}\enspace \vadjust{\penalty5000}}

\def\undertext#1{\vtop{\hbox{#1}\kern 1pt \hrule}}
\def\subsubsection#1{\par
   \ifnum\the\lastpenalty=30000\else \penalty-100\smallskip \fi
   \noindent\hbox{#1}\enspace \vadjust{\penalty5000}}

\def\appendix#1{\cleareqnumber \s@cfalse \global\advance\appnumber by1
   \message{Appendix \char\the\appnumber: #1}
   \par \ifnum\the\lastpenalty=30000\else
   \penalty-200\vskip\sectionskip \spacecheck\sectionminspace\fi
   \noindent {\caps\enspace Appendix \char\the\appnumber\quad #1}\par
   \nobreak\vskip\headskip \penalty 30000 }
\clearsecnumber
\clearappnumber
%
% macros for references, acknowledgements, and note added
%
\def\ack{\par\penalty-100\medskip \spacecheck\sectionminspace
   \line{\iftwelv@\fourteencp\else\twelvecp\fi\hfil ACKNOWLEDGEMENTS\hfil}%
\nobreak\vskip\headskip }
\def\refs{\begingroup \par\penalty-100\medskip \spacecheck\sectionminspace
   \line{\iftwelv@\fourteencp\else\twelvecp\fi\hfil REFERENCES\hfil}%
\nobreak\vskip\headskip \frenchspacing }
\def\endrefs{\par\endgroup}
%--- Note added
%
\newcount\refnumber
\def\clearrefnumber{\refnumber=0}  \clearrefnumber
\newwrite\R@fs                              %This opens a file .refs with
\immediate\openout\R@fs=\jobname.references %the references in order of
                                            %appearance.
\def\closerefs{\immediate\closeout\R@fs} %close file so that TeX can read it
\def\refsout{\closerefs\refs
\catcode`\@=11                          % we must do this since the
\input\jobname.references               % references expand to
\catcode`\@=12			        % primitives containing @'s
\endrefs}
\def\refitem#1{\item{{\bf #1}}}%just bolds it so that \bf does not expand
\def\ifundefined#1{\expandafter\ifx\csname#1\endcsname\relax}
%
%  new reference macros.  Now just say \[_label], and in the
%  ref_defs.tex file type \refdef[_label]{reference}
%
\def\[#1]{\ifundefined{#1R@FNO}%
\global\advance\refnumber by1%
\expandafter\xdef\csname#1R@FNO\endcsname{[\the\refnumber]}%
\immediate\write\R@fs{\noexpand\refitem{\csname#1R@FNO\endcsname}%
\noexpand\csname#1R@F\endcsname}\fi{\bf \csname#1R@FNO\endcsname}}
\def\refdef[#1]#2{\expandafter\gdef\csname#1R@F\endcsname{{#2}}}
%
% equation numbering macros
%
%  better than before.  just do \(_label) both to refer or to
%                         define.
%
%        the generic equation is \()
%
%
%
\newcount\eqnumber
\def\cleareqnumber{\eqnumber=0}
\newif\ifal@gn \al@gnfalse  % this is true if within an \eqalignno
% at some point try the following:
%\def\eqnalign#1{\al@gntrue \vbox{\eqalignno{#1}} \al@gnfalse}
% but meanwhile let`s define a new macro...
\def\veqnalign#1{\al@gntrue \vbox{\eqalignno{#1}} \al@gnfalse}
\def\eqnalign#1{\al@gntrue \eqalignno{#1} \al@gnfalse}
\def\(#1){\relax%
\ifundefined{#1@Q}
 \global\advance\eqnumber by1
 \ifs@cd
  \ifs@c
   \expandafter\xdef\csname#1@Q\endcsname{{%
\noexpand\rm(\the\secnumber .\the\eqnumber)}}
  \else
   \expandafter\xdef\csname#1@Q\endcsname{{%
\noexpand\rm(\char\the\appnumber .\the\eqnumber)}}
  \fi
 \else
  \expandafter\xdef\csname#1@Q\endcsname{{\noexpand\rm(\the\eqnumber)}}
 \fi
 \ifal@gn
    & \csname#1@Q\endcsname
 \else
    \eqno \csname#1@Q\endcsname
 \fi
\else%
\csname#1@Q\endcsname\fi\global\let\@Q=\relax}
%
% macros for running heads and page numbering
%
\newif\iffrontpage \frontpagefalse
\headline={\hfil}
\footline={\iffrontpage\hfil\else \hss\twelverm
-- \folio\ --\hss \fi }
\def\monthname{\relax\ifcase\month 0/\or January\or February\or
   March\or April\or May\or June\or July\or August\or September\or
   October\or November\or December\else\number\month/\fi}
\hsize=14 truecm
\vsize=22 truecm
\skip\footins=\bigskipamount
\normalspace
%
%%%%%%%%%%%%%%%%%%%%%%%%%%%%%%%%%%%%%%%%%%%%%%%%%%%%%%%%%%%%%%%%%%%%%%%
%
%   Here come macros for title pages.
%
\newskip\frontpageskip
\newif\ifp@bblock \p@bblocktrue
\newif\ifm@nth \m@nthtrue
\newtoks\pubnum
\newtoks\pubtype
\newtoks\m@nthn@me
\newcount\Ye@r
\advance\Ye@r by \year
\advance\Ye@r by -1900
\def\Year#1{\Ye@r=#1}%--- set the year by hand
\def\Month#1{\m@nthfalse \m@nthn@me={#1}}
\def\m@nthname{\ifm@nth\monthname\else\the\m@nthn@me\fi}
\def\titlepage{\global\frontpagetrue\hrule height\z@ \relax
               \ifp@bblock\pubblock\fi\relax }
\def\endtitlepage{\vfil\break
                  \frontpagefalse} %I took a \pageno=1 from here
\def\bonntitlepage{\global\frontpagetrue\hrule height\z@ \relax
               \ifp@bblock\pubblock\fi\relax }
\frontpageskip=12pt plus .5fil minus 2pt
\pubtype={\iftwelv@\twelvesl\else\tensl\fi\ (Preliminary Version)}
\pubnum={?}
\def\nopubblock{\p@bblockfalse}
\def\pubblock{\line{\hfil\iftwelv@\twelverm\else\tenrm\fi%
BONN--HE--\number\Ye@r--\the\pubnum\the\pubtype}
              \line{\hfil\iftwelv@\twelverm\else\tenrm\fi%
\m@nthname\ \number\year}}
\def\title#1{\vskip\frontpageskip\Titlestyle{\caps #1}\vskip3\headskip}
%                                 ^---notice capital Titlestyle...
\def\author#1{\vskip.5\frontpageskip\titlestyle{\caps #1}\nobreak}

\def\authors{\vskip\frontpageskip\noindent}
\def\address#1{\par\kern 5pt\titlestyle{%\iftwelv@\twelvepoint\else\tenpoint\fi
\it #1}}
\def\andaddress{\par\kern 5pt \centerline{\sl and} \address}
\def\addresses{\vskip\frontpageskip\noindent\interlinepenalty=9999}
\def\abstract#1{\par\dimen@=\prevdepth \hrule height\z@ \prevdepth=\dimen@
   \vskip\frontpageskip\spacecheck\sectionminspace
   \centerline{\iftwelv@\fourteencp\else\twelvecp\fi ABSTRACT}\vskip\headskip
   {\noindent #1}}
%

%
%%%%%%%%%%%%%%%%%%%%%%%%%%%%%%%%%%%%%%%%%%%%%%%%%%%%%%%%%%%%%%%%%%
%
% macros for leaders, boxes, underline, ...
%
\def\leaderfill{\leaders\hbox to 1em{\hss.\hss}\hfill}%--- leading ...
 %--- underline
\def\boxit#1{\vcenter{\hrule\hbox{\vrule\kern8pt
      \vbox{\kern8pt#1\kern8pt}\kern8pt\vrule}\hrule}}%--- box
 %--- box in $$...$$

%
%%%%%%%%%%%%%%%%%%%%%%%%%%%%%%%%%%%%%%%%%%%%%%%%%%%%%%%%%%%%%%%%%%
%
%  Now come basic non-math macros
%
\def\ref#1{{\bf [#1]}}%--- [ref]
%--- et al.
\def\ie{{\it i.e.\/}}%--- i.e.
%--- e.g.
%--- Cf.
%--- cf.
 %--- double left quote
\def\th{{\rm th}}%--- th as in fifth
\def\nl{\hfil\break}%--- new line
%--- just an abbrev.
%--- 1/2
%
%%%%%%%%%%%%%%%%%%%%%%%%%%%%%%%%%%%%%%%%%%%%%%%%%%%%%%%%%%%%%%%%
%
% Now some math macros
%
%%%%%%%%%%%%%%%%%%%%%%%%%%%%%%%%%%%%%
%
% First macros for theorems, definitions, ...
%
\newif\ifm@thstyle \m@thstylefalse
\def\mathstyle{\m@thstyletrue}
\def\proclaim#1#2\par{\smallbreak\begingroup%        small --> med???
\advance\baselineskip by -0.25\baselineskip%
\advance\belowdisplayskip by -0.35\belowdisplayskip%
\advance\abovedisplayskip by -0.35\abovedisplayskip%
    \noindent{\caps#1.\enspace}{#2}\par\endgroup%
\smallbreak}%--- defs, thms, ...                     small --> med???
\def\m@kem@th<#1>#2#3{%
\ifm@thstyle \global\advance\eqnumber by1
 \ifs@cd
  \ifs@c
   \expandafter\xdef\csname#1\endcsname{{%
\noexpand #2\ \the\secnumber .\the\eqnumber}}
  \else
   \expandafter\xdef\csname#1\endcsname{{%
\noexpand #2\ \char\the\appnumber .\the\eqnumber}}
  \fi
 \else
  \expandafter\xdef\csname#1\endcsname{{\noexpand #2\ \the\eqnumber}}
 \fi
 \proclaim{\csname#1\endcsname}{#3}
\else
 \proclaim{#2}{#3}
\fi}
%
%
%  To use the new Math macros...
%
%         \Thm<_label>{Statemenet of the Thm} etc...
%
%     where _label is the label of the Thm.  The generic
%     Thm has an empty label.
%
%     To refer to it, just say \<_label>
%
\def\Thm<#1>#2{\m@kem@th<#1M@TH>{Theorem}{\sl#2}}%--- Theorem
\def\Prop<#1>#2{\m@kem@th<#1M@TH>{Proposition}{\sl#2}}%--- Proposition
\def\Def<#1>#2{\m@kem@th<#1M@TH>{Definition}{\rm#2}}%--- Definition
\def\Lem<#1>#2{\m@kem@th<#1M@TH>{Lemma}{\sl#2}}%--- Lemma
\def\Cor<#1>#2{\m@kem@th<#1M@TH>{Corollary}{\sl#2}}%--- Corollary
\def\Conj<#1>#2{\m@kem@th<#1M@TH>{Conjecture}{\sl#2}}%--- Conjecture
\def\Rmk<#1>#2{\m@kem@th<#1M@TH>{Remark}{\rm#2}}%--- Remark
\def\Exm<#1>#2{\m@kem@th<#1M@TH>{Example}{\rm#2}}%--- Example
\def\Qry<#1>#2{\m@kem@th<#1M@TH>{Query}{\it#2}}%--- Query
%
%--- Proof
%

\def\<#1>{\csname#1M@TH\endcsname}
%
% We then continue with basic mathematics
%
%--- def over =
%--- Halmos Q.E.D.

%--- implies
%--- is implied by
%--- if and only if
\def\lapprox{\hbox{\lower3pt\hbox{$\buildrel<\over\sim$}}}% approx lt
\def\gapprox{\hbox{\lower3pt\hbox{$\buildrel<\over\sim$}}}% approx gt
\def\quotient#1#2{#1/\lower0pt\hbox{${#2}$}}%--- factor objects
%
% Arrow stuff
%
%--- injective map
%--- surjective map
%--- bijective map
\def\to{\rightarrow}%--- mapping
%--- long mapping
%--- isom over -->
%--- just an abbrev.
%

%
 %--- commutative diagram macro
%-- lin map over arrow
 %--- map in complex
%
% Numbers...
%
 %--- reals
 %--- complex nos.
 %--- quaternions
 %--- integers
 %--- rationals
\def\nats{{\bf N}} %--- naturals
 %--- ground field
%
% Algebra
%
%--- Hom(omorphisms)
%--- tr(ace)
\def\Tr{\mathop{\rm Tr}}%--- Tr(ace)
%--- End(omorphisms)
%--- Mor(phisms)
%--- Aut(omorphisms)
%--- aut(omorphisms)
%--- supertrace
%--- superdeterminant
\def\ker{\mathop{\rm ker}}%--- kernel
%--- cokernel
%--- image
\def\underrightarrow#1{\vtop{\ialign{##\crcr
      $\hfil\displaystyle{#1}\hfil$\crcr
      \noalign{\kern-\p@\nointerlineskip}
      \rightarrowfill\crcr}}} %--- modification of \overrightarrow
\def\underleftarrow#1{\vtop{\ialign{##\crcr
      $\hfil\displaystyle{#1}\hfil$\crcr
      \noalign{\kern-\p@\nointerlineskip}
      \leftarrowfill\crcr}}}  %--- modification of \overleftarrow

%
% Brackets,...
%
\def\comm#1#2{\left[#1\, ,\,#2\right]}%--- [ , ]
%--- { , }
%--- [ , }
%--- structure const.
%
% Analysis anyone?
%
%--- Lie derivative
% lft var derivative
% rgt var derivative
\def\vder#1#2{{{{\delta}{#1}}\over{{\delta}{#2}}}}%--- vartnl derivative
% double vartl deriv.
%--- rgt prtl derivative
\def\pder#1#2{{{\partial #1}\over{\partial #2}}}%--- partial derivative
%
%                                                           dble partl deriv.
%--- full lft derivative
%--- full rgt derivative
%--- full derivative
%--- laplacian
%--- cov. ext. der.
%
% Dirac slashes
%
%--- D slash
%--- del slash
%--- A slash
%
%%%%%%%%%%%%%%%%%%%%%%%%%%%%%%%%%%%%%%%%%%%%%%%%%%%%%%%%%%%%%%%%
%
%  These are the macros to make Young tableaux
%
\newdimen\unit
\newdimen\redunit
%
%   this puts the ref. point of #1 at coordinates (#2,#3)
%
\def\p@int#1:#2 #3 {\rlap{\kern#2\unit
     \raise#3\unit\hbox{#1}}}
%
% this defines the sides of the tableau
% notice that \rver and \lver coincide
% It would have been natural for \rver to have negative
% width but that does not print in TeX...
%
\def\th@r{\vrule height0\unit depth.1\unit width1\unit}
\def\bh@r{\vrule height.1\unit depth0\unit width1\unit}
\def\lv@r{\vrule height1\unit depth0\unit width.1\unit}
\def\rv@r{\vrule height1\unit depth0\unit width.1\unit}
%
% this is the tableau: the .9 is due to the unnatural definition
% of \rver
%
\def\t@ble@u{\hbox{\p@int\bh@r:0 0
                   \p@int\lv@r:0 0
                   \p@int\rv@r:.9 0
                   \p@int\th@r:0 1
                   }
             }
%
% we now define the tableau at a particular location
%
\def\t@bleau#1#2{\rlap{\kern#1\redunit
     \raise#2\redunit\t@ble@u}}
%
%  Now a macro to make a column of #1 tableaux down at (#2,#3)
%
\newcount\n
\newcount\m
\def\makecol#1#2#3{\n=0 \m=#3
  \loop\ifnum\n<#1{}\advance\m by -1 \t@bleau{#2}{\number\m}\advance\n by 1
\repeat}
%
%   Now a macro to make a row of #1 tableaux at (#2,#3) to the right
%
\def\makerow#1#2#3{\n=0 \m=#3
 \loop\ifnum\n<#1{}\advance\m by 1 \t@bleau{\number\m}{#2}\advance\n by 1
\repeat}
%
% Some useful ready made Young tableaux
%
\def\checkunits{\ifinner \unit=6pt \else \unit=8pt \fi
                \redunit=0.9\unit } %these are the basic sizes
\def\ytsym#1{\checkunits\kern-.5\unit
  \vcenter{\hbox{\makerow{#1}{0}{0}\kern#1\unit}}\kern.5em} % #1 symmetrized
%                                                             tableaux
\def\ytant#1{\checkunits\kern.5em
  \vcenter{\hbox{\makecol{#1}{0}{0}\kern1\unit}}\kern.5em} % #1 antisymmetrized
%                                                            tableaux
\def\ytwo#1#2{\checkunits
  \vcenter{\hbox{\makecol{#1}{0}{0}\makecol{#2}{1}{0}\kern2\unit}}
                  \ } % 2 column #1 #2 (left->right) Young tableau
\def\ythree#1#2#3{\checkunits
  \vcenter{\hbox{\makecol{#1}{0}{0}\makecol{#2}{1}{0}\makecol{#3}{2}{0}%
\kern3\unit}}
                  \ } % 3 column #1 #2 #3 (left->right) Young tableau
%
%%%%%%%%%%%%%%%%%%%%%%%%%%%%%%%%%%%%%%%%%%%%%%%%%%%%%%%%%%%%%%%%%%%%%
%
% Finally some useful macros for journals, ...
%

\def\NPB#1#2#3{{\sl Nucl. Phys.} {\bf B#1} (#2) #3}

\def\PLB#1#2#3{{\sl Phys. Lett.} {\bf #1B} (#2) #3}

\def\AdM#1#2#3{{\sl Advances in Math.} {\bf #1} (#2) #3}

\catcode`\@=12 % @ no longer a letter
%
%
%
%   These are the local macros for not-kp note.
%
\def\d{\partial}
\let\pb=\anticomm

% kth partial
\def\vbar{\overline{v}}

\def\D{\nabla}
\def\gU{{\cal U}}
\def\gV{{\cal V}}
\def\chq{\mathop{{\rm ch}}\nolimits_q}
\def\phibar{\overline{\phi}}
\def\fr#1/#2{\hbox{${#1}\over{#2}$}}
\refdef[Dickey]{L.~A.~Dickey,  {\sl Soliton equations and Hamiltonian
systems},  Advanced Series in Mathematical Physics Vol.12,  World
Scientific Publ.~Co..}
\refdef[Depir]{D.~A.~Depireux, Preprint LAVAL PHY-21-92, {\tt
hepth@xxx/9203062}.}
\refdef[AFGZ]{H.~Aratyn, L.~A.~Ferreira, J.~F.~Gomes, and
A.~H.~Zimerman, Preprint IFT-P/020/92-SAO-PAULO, {\tt
hep-th/9206096}.}
\refdef[YuWu]{F.~Yu and Y.-S.~Wu, Preprint UU-HEP-91/19, {\tt
hepth@xxx/9112009}.}
\refdef[WKP]{L.~A.~Dickey, {\sl Annals of the New York Academy of
Science} {\bf 491}(1987) 131.;\nl J.~M.~Figueroa-O'Farrill, J.~Mas,
and E.~Ramos, \PLB{266}{1991}{298};\nl F.~Yu and Y.-S.~Wu,
\NPB{373}{1992}{713}.}
\refdef[Mulase]{M.~Mulase, \AdM{53}{1984}{57}.}
\refdef[Andrews]{G.~E~.Andrews, {\sl The Theory of Partitions},
Addison-Wesley~Pub.~Co. 1976.}
\overfullrule=0pt
\unsectioned
\def\pubblock{ \line{\hfil\twelverm BONN--HE--92--17}
	       \line{\hfil\twelverm US--FT--4/92}
               \line{\hfil\twelverm KUL--TF--92/26}
               \line{\hfil\twelvett hep-th/9207013}
               \line{\hfil\twelverm July 1992}}
%
%\nopubblock
\titlepage
\title{ON THE TWO-BOSON PICTURE OF THE KP HIERARCHY}
\authors
{\caps Jos\'e~M.~Figueroa-O'Farrill$^1$\footnote{$^\flat$}{\tt
e-mail: figueroa@pib1.physik.uni-bonn.de.}},
{\caps Javier~Mas$^2$\footnote{$^\natural$}{\tt e-mail:
jamas@gaes.usc.es}}, and {\caps
Eduardo~Ramos$^3$\footnote{$^\sharp$}{{\tt e-mail:
fgbda06@blekul11.bitnet.} {\rm Address after October 1992: Queen Mary
and Westfield College, UK}}}
\addresses
$^1${\it Physikalisches Institut der Universit\"at Bonn, Nu{\ss}allee
12, W-5300 Bonn 1, GERMANY}\hfil\break\noindent
$^2${\it Departamento de F{\'\i}sica de Part{\'\i}culas Elementales,
Universidad de Santiago, E-15706 Santiago de Compostela,
SPAIN}\hfil\break\noindent
$^3${\it Instituut~voor~Theoretische~Fysica, Universiteit~Leuven,
Celestijnenlaan 200D, B-3001~Heverlee, BELGIUM}
\abstract{A two-boson realization of the second hamiltonian structure
for the KP hierarchy has recently appeared in the literature.
Furthermore, it has been claimed that this is also a realization of
the hierarchy itself.  This is surprising because it would mean that
the dynamics of the KP hierarchy---which in its usual formulation
requires an infinite number of fields---can be described with only
two.  The purpose of this short note is to point out the almost
obvious fact that the hierarchy described by the two bosons is not the
KP hierarchy but rather a reduction thereof---one which is moreover
incompatible with the reduction to the KdV-type subhierarchies.}
\endtitlepage
\section{Introduction}

The KP hierarchy\fnote{Brevity forbids that we dwell on notation or on
formalism, for both of which we direct the reader to Dickey's
comprehensive treatise \[Dickey].} is the isospectral problem of the
one-dimensional pseudodifferential operator given by
$$\Lambda = \d + \sum_{j\geq 0} u_j \d^{-j}~,\(KPop)$$
where the $\{u_j\}$ are assumed independent generators of a
differential ring.  The isospectral deformations are of Lax type and
given by
$$\pder{\Lambda}{t_\ell} = \comm{\Lambda^\ell_+}{\Lambda}~,\qquad \ell
\in\nats ~.\(KPflows)$$
These flows are easily shown to commute and to be hamiltonian relative
to an infinite number of (Dickey--Radul) bihamiltonian structures with
hamiltonian functions (\ie, conserved charges) given by the Adler
trace of the powers of the KP operator \(KPop): $H_\ell =
{1\over\ell} \Tr \Lambda^\ell$.  The simplest and the original of
these hamiltonian structures is given by
$$\pb{F}{G} = \Tr \left[ (\Lambda\,dF)_+\Lambda\,dG - (dF\,\Lambda)_+
dG\,\Lambda\right] ~,\(WKP)$$
where $dF = \sum_{j\geq0} \d^{j-1}\vder{F}{u_j}$ and similarly for
$dG$.

The KP hierarchy was originally introduced in order to unify the
treatment of the generalized KdV hierarchies.  The $n^\th$-order KdV
hierarchy ($n$-KdV, for short) is the hierarchy of isospectral flows
of the differential operator
$$L = \d^n + \sum_{j=0}^{n-1} w_j\d^j~,\(nKdVop)$$
where, again, one assumes that the $\{w_j\}$ are independent and
generate freely a differential ring.  By taking its $n^\th$-root, we
can think of the space of such operators as the subspace of KP
operators \(KPop) satisfying the constraint $\Lambda^n_- = 0$.  This
constraint is preserved by the Lax flows and the constrained hierarchy
is precisely $n$-KdV.  Moreover, the $n$-KdV hierarchy is hamiltonian
relative to the Adler--Gel'fand--Dickey bihamiltonian structure, which
extends naturally (for each $n$) to one of the Dickey--Radul
bihamiltonian structures of the KP hierarchy.

The ``second'' bracket of this bihamiltonian structure has an
expression analogous to \(WKP).  If $F$ and $G$ are functions of $L$,
their Poisson bracket is given by
$$\pb{F}{G} = \Tr \left[ (L\,dF)_+L\,dG - (dF\,L)_+
dG\,L\right] ~,\(GDbra)$$
where now $dF = \sum_{j=0}^{n-1} \d^{-j-1}\vder{F}{w_j}$ and similarly
for $dG$.  The Kupershmidt--Wilson theorem relates this bracket with a
much simpler one in another set of variable to which the $\{w_j\}$ are
related by the Miura transformation.  Indeed, formally factorizing the
operator \(nKdVop) into linear terms
$$L = (\d-v_1)(\d-v_2)\cdots (\d -v_n)~,\(Miura)$$
defines an embedding of the differential ring generated by the
$\{w_j\}$ into the differential ring generated by the $\{v_j\}$.
On the space of the $\{v_j\}$ we can define a very simple bracket,
namely
$$\pb{v_i(z)}{v_j(w)} = \delta_{ij} \delta'(z-w)~,\(miurapb)$$
which induces a bracket on the $\{w_j\}$ through their embedding via
the Miura transformation.  What the Kupershmidt--Wilson theorem
asserts is that the induced bracket is precisely \(GDbra).

Although the Kupershmidt--Wilson theorem holds for $n$-KdV for all
$n$, a similar result of the KP hierarchy is lacking, despite recent
claims to the contrary in the literature \[Depir], \[AFGZ].  The
origin of these claims is the remarkable transformation due to Wu and
Yu of the KP operator \(KPop) in terms of two bosons \[YuWu], obtained
by deforming in a natural way the bosonized two-fermion realization of
$W_\infty$.  Let us consider the following ``factorization'' of the KP
operator\fnote{This is a slight modification of the original
transformation of Wu and Yu in that we add a zeroth order term to the
KP operator.  We find this factorization has nicer formal properties.
Of course, the main points of this discussion are impervious to this
modification.}
$$\Lambda = \D + \vbar \D^{-1} v~,\(WuYutransf)$$
where $\D = \d + \vbar + v$.  This maps the differential ring
generated by the $\{u_j\}$ to the
differential ring generated by $v$ and $\vbar$ in such a way that, if
on this latter ring we define the Poisson brackets
$$\veqnalign{\pb{v(z)}{\vbar(w)} &= \delta'(z-w)\cr
\pb{v(z)}{v(w)} &= \pb{\vbar(z)}{\vbar(w)} = 0~,\cr}$$
the induced brackets coincide with the ``second'' hamiltonian
structure for the KP hierarchy \[WKP].  This means that the KP flows
\(KPflows) induce flows in $v$ and $\vbar$ which are simply given by
$$\veqnalign{\pder{v}{t_\ell} &= \left(
\pder{H_\ell}{\vbar}\right)'\cr
\pder{\vbar}{t_\ell} &=
\left(\pder{H_\ell}{v}\right)'~,\(bosonflows)\cr}$$
where the hamiltonians are now expressed in terms of $v$ and $\vbar$
via \(WuYutransf).

What has been claimed in \[Depir] and \[AFGZ] is that this hierarchy
is in fact the KP hierarchy.  This is not quite correct.  We will
prove that, unlike the Miura transformation \(Miura), the mapping
\(WuYutransf) is not an embedding.  In other words, the mapping $u_j
\mapsto u_j(v,\vbar)$ induces relations on the $\{u_j\}$ which were
not present originally.  In the KP hierarchy the coefficients
$\{u_j\}$ of the KP operator \(KPop) are assumed to be independent
and, in fact, the initial value problem for the KP hierarchy of
equations is well-defined \[Mulase] (within a given class of
functions) by specifying $w_j(x,t_\ell=0)$ independently for all $j$.
Because of the extra relations imposed by \(WuYutransf), it is clear
that there is initial value data for the KP hierarchy which can never
be described by the two-boson hierarchy.  Basically there is a large
portion of the ``phase space'' of the KP hierarchy that the two-boson
hierarchy does not describe.

To prove this we will simply show that the map between differential rings
induced by the transformation \(WuYutransf) is not an embedding.  We
do this by computing the asymptotic dimensions of the differential
rings in question.  We make this manifest by explicitly writing down
the simplest of the relations between the $\{u_i\}$ and showing
explicitly that further reducing to the KdV hierarchy essentially
collapses the phase space.
\section{The Miura-like transformation of Wu and Yu}

Let $\gU$ and $\gV$ denote, respectively, the differential rings
generated by the $\{u_i\}$ and by $\vbar, v$.  The Wu--Yu
transformation defined by \(WuYutransf) induces a differential ring
map $\varphi : \gU \to \gV$ which can be explicitly described as
follows.  Let us introduce the formal integrals $\phibar,\phi$ of
$\vbar,v$: $\phi' = v$ and $\phibar' =\vbar$.  Then we can write
$$\D = \d + \vbar + v = e^{-\phibar - \phi}\,\d\,e^{\phibar +
\phi}~,\()$$
and therefore
$$\Lambda = e^{-\phibar - \phi} \left[ \d + \vbar \d^{-1} v\right]
e^{\phibar + \phi}~.\()$$
Using the generalized Leibniz rule for $\d^{-1}$, we obtain
$$\Lambda = \D + \sum_{j=0}^\infty (-1)^j \vbar (\D^j \cdot
v)\d^{-j-1}~,\()$$
from where we can read off the action of the map $\varphi$ on the
generators:
$$\varphi(u_0) = \vbar + v\qquad \varphi(u_{i>0}) = \vbar
(-\D)^{i-1}\cdot v~.\(phiongens)$$
This extends uniquely to a map of differential rings in the obvious
way.

The differential rings $\gU$ and $\gV$ have natural gradations.  We can
define the following degrees: $[\vbar]=[v]=1$, $[u_j]=j+1$, and
$[f']=[f]+1$ for any homogeneous element $f$.  If we define $\gU_n$
(respectively $\gV_n$) as the subspace of $\gU$ (respectively $\gV$)
of homogeneous elements of degree $n$, we find that $\gU =
\bigoplus_{n=0}^\infty \gU_n$ and $\gV = \bigoplus_{n=0}^\infty \gV_n$
and, moreover, that the map $\varphi$ respects the grading and thus
induces linear maps $\varphi_n : \gU_n \to \gV_n$.  It is clear that
$\varphi$ would be one-to-one if and only if each of the $\varphi_n$
would be too.  We will show, however, that for $n$ large enough, this
cannot be the case since the dimension of $\gU_n$ exceeds that of
$\gV_n$.

To see this, let us introduce the following formal sums:
$$\veqnalign{\chq\gU &\equiv \sum_{n=0}^\infty \dim\gU_n\, q^n = 1 +
\sum_{n=1}^\infty U(n) q^n\cr
\chq\gV &\equiv \sum_{n=0}^\infty \dim\gV_n\, q^n = 1 +
\sum_{n=1}^\infty V(n) q^n~,\()\cr}$$
which defines $U(n)$ and $V(n)$.  We can compute these sums exactly by
noticing that $\gU$ and $\gV$ are polynomial rings in their
generators and their derivatives.  Indeed, $\gU$ is the polynomial
ring in the variables $u_i^{(j)}$ for $i,j=0,1,2,\ldots$ of degree
$[u_i^{(j)}] = i+j+1$.  Therefore,
$$\chq\gU = \prod_{i=0}^\infty \prod_{j=0}^\infty {1\over
(1-q^{i+j+1})} = \prod_{n=1}^\infty {1\over (1-q^n)^n}~.\(Uchar)$$
Similarly, $\gV$ is a polynomial ring in $v^{(i)}$ and $\vbar^{(i)}$
for $i=0,1,2,\ldots$ and its formal character is thus readily computed
to be
$$\chq\gV = \prod_{n=1}^\infty {1\over (1-q^n)^2}~.\(Vchar)$$
We are interested in computing the asymptotic dimensions $U(n)$ and
$V(n)$.  This follows from a simple application of the theorem of
Meinardus on the asymptotics of infinite produce generating functions
(see, for example, Chapter 6 in \[Andrews]).  After a brief
inspection we find
$$\veqnalign{U(n) &\sim A n^{-5/6} \exp \left(B n^{2/3}\right)~,\cr
\noalign{\hbox{and}}
V(n) &\sim C n^{-5/4} \exp \left(D n^{1/2}\right)~,\cr}$$
for some constants $A,B,C,D$ which need not concern us here.  Notice,
that asymptotically $U(n) > V(n)$ which proves that the maps
$\varphi_n$ cannot be injective after some $n$ and, therefore, that
$\varphi$ is not an embedding of differential rings.  Its kernel will
be a differential ideal of $\gV$ generated by some polynomial
relations among the $\{u_i\}$.  It would be very interesting to be
able to write down these relations in general and thus obtain some
intuition about the subhierarchy of KP defined by the two bosons.  We
have thus far only been able to compute the simplest ones.
\section{An explicit polynomial relation and inconsistency of the KdV
reduction}

We just saw that for $n$ large enough we expect to find polynomial
relations among the $\{u_i\}$.  In fact, one need not go very far to
find them.  A short calculation yields
$$\veqnalign{\chq\gU ={}& 1 + q+ 3q^2 + 6q^3 + 13q^4 + 24q^5 + 48q^6 +
86q^7\cr
& + 160q^8 + 282q^9 + 500q^{10} + O(q^{11})\cr
\chq\gV ={}& 1 + 2q+ 5q^2 + 10q^3 + 20q^4 + 36q^5 + 65q^6 + 110q^7\cr
& + 185q^8 + 300q^9 + 481q^{10} + O(q^{11})~,\cr}$$
which shows that our first polynomial relation will occur for $n\leq
10$.

In fact, one finds that the first relations appear at $n=6$.  There
are three independent ones, of which the simplest is
$$P = u_3 u_1 - u_2^2 + u_2'u_1 - u_2u_1'~.\(polyrel)$$
One can easily check that the above differential polynomial vanishes
after substituting the explicit expressions of $u_1$, $u_2$ and $u_3$
in terms of $\vbar$ and $v$.  The other two relations involve $u_0$
and would therefore not be relations in the map defined originally by
Wu and Yu (see previous footnote).

Let's see how this relation behaves under a further reduction to the
KdV hierarchy.  This reduction corresponds to
$\left(\Lambda^2\right)_-=0$.  This imposes polynomial relations among
the $\{u_i\}$ of the form $u_{i>2} = p_i(u_{j<i})$.  In
particular---dropping $u_0$ for convenience---we find
$$u_2 = -\fr1/2  u_1' \quad \hbox{and} \quad u_3 = \fr1/4 u_1'' -
\fr1/2 u_1^2 ~.\(KdVrels)$$
We see that \(polyrel) is consistent with \(KdVrels) if and only if
$\log u_1$ obeys the Liouville-type equation
$$(\log u_1)'' = -2u_1~.\(Liouville)$$
However, further relations appear which impose other relations on
$u_1$ besides \(Liouville) and cause further collapse of the phase
space.  Notice, parenthetically, that these are all kinematical
constraints.  The dynamics impose no further constraints since both
the KdV and Wu--Yu reductions are preserved by the KP flows: for the
KdV equation the result is classical, whereas for the Wu--Yu reduction
it follows from the fact that the hamiltonian structures correspond.
\section{Conclusions}

In summary, we have shown that unlike in the case of the $n$-KdV
hierarchy, the Miura-like transformation induced by the factorization
\(WuYutransf) of the KP operator is not an embedding of differential
rings.  This has the following somewhat formal consequences.

On the one hand, the result of Wu and Yu on the induced hamiltonian
structures does not constitute---the authors never claimed it did---an
independent proof of the Jacobi identity for the bracket \(WKP), as
the Kupershmidt--Wilson theorem does for the case of the
Adler--Gel'fand--Dickey bracket.  This is because in checking the
Jacobi identity in terms of $\vbar$ and $v$ we are actually not
performing the calculation in the ring $\gU$ but in its image under
$\varphi$.  Thus a priori the Jacobi identities hold only modulo $\ker
\varphi$.  It is in this sense that a theorem analogous to the
Kupershmidt--Wilson theorem does not yet exist for the KP hierarchy.
Of course, the Jacobi identity for the bracket \(WKP) has been proven
directly in \[WKP].

And on the other hand, the two-boson hierarchy is not the KP hierarchy
but rather a reduction thereof.  The existence of polynomial relations
also make it impossible a priori to reconstruct the KP hierarchy from
this one, since there is no guarantee that all the conserved charges
will remain nontrivial when written in terms of $\vbar$ and $v$.  It
would be very interesting to elucidate this point.

Despite these formal drawbacks, \(WuYutransf) remains a most
remarkable transformation and the two-boson hierarchy an interesting
novel reduction of KP.

\ack

One of us (JMF) benefitted greatly from conversations with S.~Stanciu
and is therefore grateful.
\refsout
\bye